\documentclass[structabstract]{aa}
\usepackage{graphicx}
\usepackage{epstopdf}
\usepackage{amssymb,amsmath}
\usepackage[colorlinks]{hyperref}
\usepackage{aalongtable}
\usepackage{lscape}
\usepackage[dvipsnames]{xcolor}
\usepackage{natbib}
\usepackage{enumitem}
\usepackage{ulem}
\usepackage{listings}
\usepackage{tablefootnote}

\newcommand{\foothref}[1]{\footnote{\url{#1}}} 
\newcommand{\gr}{{$\gamma$-ray}}
\newcommand{\lsim}{{\lower.5ex\hbox{$\; \buildrel < \over \sim \;$}}}
\newcommand{\gsim}{{\lower.5ex\hbox{$\; \buildrel > \over \sim \;$}}}
\newcommand{\nup}{$\nu_{\rm peak}$}
\newcommand{\nufnu}{$\nu$f$({\nu})$}

\newcommand{\ergs}{erg cm$^{-2}$ s$^{-1}$}

\begin{document}
   \title{The Open Universe survey of Swift-XRT GRB fields: \\ a flux-limited sample of HBL blazars.}
   \author{
          P. Giommi   \inst{1,2,3}
          \and Y. L. Chang  \inst{4}
          \and S. Turriziani\inst{5}
          \and T. Glauch \inst{1}
          \and C. Leto   \inst{6,7} 
          \and F. Verrecchia  \inst{7,8} 
          \and P. Padovani \inst{9}
          \and A. V. Penacchioni\inst{10,15}
          \and F. Arneodo \inst{11}
          \and U. Barres de Almeida \inst{12,3}
          \and C.H. Brandt  \inst{13}
          \and M. Capalbi \inst{14}
          \and O. Civitarese\inst{10,15}
          \and V. D'Elia \inst{6,7,8} 
          \and A. Di Giovanni \inst{11}
          \and M. De Angelis\inst{6}
          \and J. Del Rio Vera\inst{16}
          \and S. Di Pippo\inst{16}
          \and R. Middei \inst{7,8}
          \and M. Perri   \inst{7,8} 
          \and A.M.T. Pollock \inst{17}
          \and S. Puccetti \inst{6}
          \and N. Ricard\inst{16}
	      \and R. Ruffini \inst{3}
	      \and N. Sahakyan\inst{18, 3}
         }        
         \institute{
          Institute for Advanced Study, Technische Universit{\"a}t M{\"u}nchen, Lichtenbergstrasse 2a, D-85748 Garching bei M\"unchen, Germany
         \and Associated to the Italian Space Agency, ASI, via del Politecnico snc, 00133 Roma, Italy
         \and ICRANet, P.zza della Repubblica 10, 65122, Pescara, Italy
         \and Tsung-Dao Lee Institute, Shanghai Jiao Tong University, 800 Dongchuan RD. Minhang District, Shanghai, China
         \and  Physics Department, Gubkin Russian State University (National Research University), 65 Leninsky Prospekt, Moscow, 119991, Russian Federation
         \and Italian Space Agency, ASI, via del Politecnico snc, 00133 Roma, Italy
         \and Space Science Data Center, SSDC, ASI, via del Politecnico snc, 00133 Roma, Italy
         \and INAF - Osservatorio Astronomico di Roma, via Frascati 33, I-00040 Monteporzio Catone, Italy
         \and European Southern Observatory, Karl-Schwarzschild-Str. 2, D-85748 Garching bei M\"unchen, Germany
         \and Institute of Physics. IFLP-CONICET. diag 113 e/63-64.(1900) La Plata. Argentina
          \and New York University Abu Dhabi, Abu Dhabi, UAE
          \and Centro Brasileiro de Pesquisas F\'isicas, Rua Dr. Xavier Sigaud 150, 22290-180, Rio de Janeiro, Brazil
         \and Jacobs University, Physics and Earth Sciences, Campus Ring 1, 28759, Bremen, Germany  
         \and INAF - Istituto di Astrofisica Spaziale e Fisica Cosmica di Palermo, via Ugo La Malfa 153, I-90146 Palermo, Italy
         \and Department of Physics. University of La Plata. 49 and 115.C.C.67 (1900) La Plata, Argentina
         \and United Nations Office for Outer Space Affairs, UNOOSA, Vienna, Austria
         \and Department of Physics and Astronomy, University of Sheffield, Hounsfield Road, Sheffield S3 7RH, England
         \and ICRANet-Armenia, Marshall Baghramian Avenue 24a, Yerevan 0019, Armenia.
         \\
         \email{giommipaolo@gmail.com}
        }
\abstract
{}
{
We have analysed all the X-ray images centred on Gamma Ray Bursts generated by Swift over the last 15 years using automatic tools that 
do not require any expertise in X-ray astronomy, producing results in excellent agreement with previous findings.
This work, besides presenting the largest medium-deep survey of the X-ray sky and a complete sample of blazars, wishes to be a step in the direction of achieving the 
ultimate goal of the Open Universe Initiative, that is to enable non expert people to fully benefit of space science data, possibly 
extending the potential for scientific discovery, currently confined within a small number of highly specialised teams, to a much larger population. 
} 
{
We have used the Swift\_deepsky Docker container encapsulated pipeline to build the largest existing flux-limited and unbiased sample of serendipitous X-ray sources. 
Swift\_deepsky runs on any laptop or desktop computer with a modern operating system. The tool automatically downloads the data and the calibration files from the archives, 
runs the official Swift analysis software and produces a number of results including images, the list of detected sources, X-ray fluxes, SED data, and spectral slope estimations.
} 
{
We used our source list to build the LogN-LogS of extra-galactic sources, which perfectly matches that estimated by other satellites.
Combining our survey with multi-frequency data we selected a complete radio flux-density limited sample of High Energy Peaked (HBL) blazars.
The LogN-LogS built with this data-set confirms that previous samples are incomplete below $\sim 20$ mJy.
}
{}
\titlerunning{Swift-XRT survey of GRB fields}
\authorrunning{Giommi, P. et al.}
\maketitle
%
\section{Introduction}\label{intro}

X-ray sky surveys have been playing a major role in astrophysics
ever since the early days of X-ray astronomy \citep[e.g.][]{Giacconi1979}.
Outside the Galactic plane the main population of X-ray sources is that of Active Galactic Nuclei \citep[AGN,][]{BH}, both jetted and non-jetted 
\citep{Padovani2017}, 
reflecting the fact that X-rays trace both the accretion onto super-massive black holes, and the radiation output of relativistic jets. 
In this paper, which follows previous similar works by \cite{Puccetti2011} and \cite{Dai2015}, we describe a serendipitous survey based on X-ray images taken when the Neil Gehrels Swift Observatory \citep[][hereafter Swift]{swift} was pointing at gamma-ray bursts (GRBs) during its first 15 years of operations. Besides being based on the largest available data set, the main peculiarity of this survey is that it has been generated and cleaned in an automatic way, without any visual or manual intervention. This was done using the Swift\_deepsky Docker pipeline, an innovative analysis tool developed in the context of the Open Universe initiative \citep[][hereafter Paper I]{paper1}, that greatly simplifies X-ray image analysis and can be run on most personal or desktop computers, even by users with no experience in X-ray astronomy. 

Since GRBs explode at random positions in the sky, this survey, after the removal of the target GRBs, constitutes an unbiased medium-deep view of the serendipitous X-ray sky that is suitable for population studies and for the estimation of the cosmological properties of cosmic sources of different types \citep{ST2019blaz}. The main improvement of this survey compared to \cite{Puccetti2011} and \cite{Dai2015} is a significant increase in the covered area, rather than reaching higher sensitivity. This is because the amount of exposure time dedicated by Swift to GRBs was largest at the beginning of the mission,
and because the need to avoid human intervention in the flagging of spurious sources, especially in the deepest exposures close to the limits of the instrument, reduces the sensitivity to values that are somewhat above the theoretical limit.  

In the following we concentrate on blazars, a remarkable type of AGN that emits non-thermal and highly variable radiation across the entire electromagnetic spectrum, from radio waves to very high-energy \gr s \citep[see e.g.][]{Urry1995,Padovani2017}, and likely also high-energy neutrinos \citep[][]{neutrino,Dissecting,Giommi2020}. 
This unique property among extragalactic sources is due to the fact that, in addition to radiating through the process of accretion onto the central super-massive black hole, that is common to all AGN, blazars also emit powerful radiation from a narrow relativistic jet that happens to be closely aligned to the direction of the Earth \citep{Padovani2017}. 

Blazars come in different types: Flat Spectrum Radio Quasars (or FSRQs) whose optical spectrum shows broad emission lines just like normal QSOs, and BL Lacertae objects (or BL Lacs) that show only very narrow lines or a completely featureless optical spectrum. Blazars are further classified according to the shape of their spectral energy distribution (SED) into low, intermediate, and high energy peaked objects, LBL (or LSP), IBL (or ISP), and HBL (or HSP) respectively, depending on the energy where the power of their synchrotron emission peaks in their SED \citep{Padovani1995,Abdo2010}. In this paper we adopt the original HBL/IBL/LBL nomenclature.

One of the still poorly understood properties of blazars concerns their cosmological evolution, which for BL Lacs has been found to be different from that of all other types of AGNs and star forming galaxies \citep[see e.g.][]{Maccacaro1984,Sedentary,Rector2000,ST2019blaz}. Early studies \citep{Rec00,Wol01b} based on small X-ray selected samples have shown that BL Lacs, and in particular those of the HBL class display no or even negative cosmological evolution. 
This peculiar behaviour has been confirmed in radio flux limited samples of the most extreme HBLs \citep{Sedentary}.
More recently \citet{ajello2014} showed that even in the case of a $\gamma$-ray selected sample low-luminosity HBL BL Lacs show strong negative evolution. 
Despite their extreme rareness, HBL blazars play a crucial role in current and future high and very-high energy \gr\ surveys, \citep{4FGL,CTA} and, likely, in multi-messenger astrophysics \citep[][and references therein]{Giommi2020}.
For the first time, we present an X-ray survey that is large and deep enough to allow the selection of a 
statistically complete flux-limited sample of blazars of this type with radio flux-densities $\lesssim 20$ 
mJy.

This work is also a demonstrator that complex data analysis projects can in principle be carried out by non experts, one of the main goals of the United Nations Open Universe initiative.

\section{The Open Universe Initiative}

Open Universe \citep{GiommiOU} is an initiative under the auspices of the United Nations Office for Outer Space Affairs (UNOOSA) with the objective of making astronomy and space science data more openly available, easily discoverable, free of bureaucratic, administrative or technical barriers, and therefore usable by the widest possible community, from professional researchers to all people interested in space science and astronomy, including students, non-professionals and amateur scholars of the subject.
One of the main goals of Open Universe is to contribute to increase productivity of space research, and stimulate a significant acceleration towards the democratisation of space science, therefore contributing to the achievement of the United Nations Sustainable Development Goals (SDGs)\foothref{http://www.unoosa.org/oosa/oosadoc/data/documents/2018/aac.105/aac.1051175_0.html}.
Another goal is to contribute to the development of open
data and web interface requirements, so as to make space
science data more understandable and attractive (reducing for example the problem of information overload, also known as infobesity), implementing lesson learned and recommendations arising from behavioural economics findings \citep{nudge,simpler}, in a way to broaden the consultation of scientific data and to bring new students and unskilled people closer to science.

The initiative was proposed by Italy to the Committee On the Peaceful Uses of Outer Space (COPUOS) in 2016, and is now actively carried out by a number of Member States and international institutions under the coordination of the United Nations Office for Outer Space Affairs (UNOOSA). 
In line with the objectives of Open Universe we have recently started a series of activities aiming at the generation of transparent space science data products.

\section{GRBs and Swift}\label{grbsandswift}

The Swift satellite was conceived and specifically designed as a panchromatic space observatory dedicated to the observations of GRBs, from the detection of the explosion in the large field of view of its Burst Alert Telescope \citep[BAT][]{bat} operating in the hard X-ray band, to the fast and automatic follow up by means of the on-board narrow fields instruments XRT \citep{xrt} and UVOT \citep{uvot} operating in the soft/medium X-ray and in the optical/UV bands, respectively. 

GRBs are the most powerful transient sources in the Universe. They are located at cosmological distances and are detected at a rate of approximately one event per day at random positions on the celestial sphere. Assuming that GRBs radiate isotropically, their energy release in X-rays and gamma rays lies in the range $10^{51}-10^{54}$ erg. 
GRBs consist of an intense and highly-variable emission in gamma rays called prompt emission, followed by the so called afterglow phase, a long-lasting activity in which the observed flux decreases with time and the emission energy shifts to lower values (X-rays, optical, IR and radio bands). The prompt phase usually lasts from milliseconds to minutes, while the afterglow duration can be from hours to weeks. There are no two GRBs detected so far with identical light-curves. The prompt emission is non-repeating, non periodic, highly variable and very energetic. 
 
The study of GRBs and the modelling of their progenitors and emission mechanisms is possible thanks to the many space and ground-based observatories operating in different energy bands that have provided or currently provide large amounts of data, part of which can also be used for other purposes, as in this paper.

\begin{figure}[ht!]
\centering
\includegraphics[width=9.0cm]{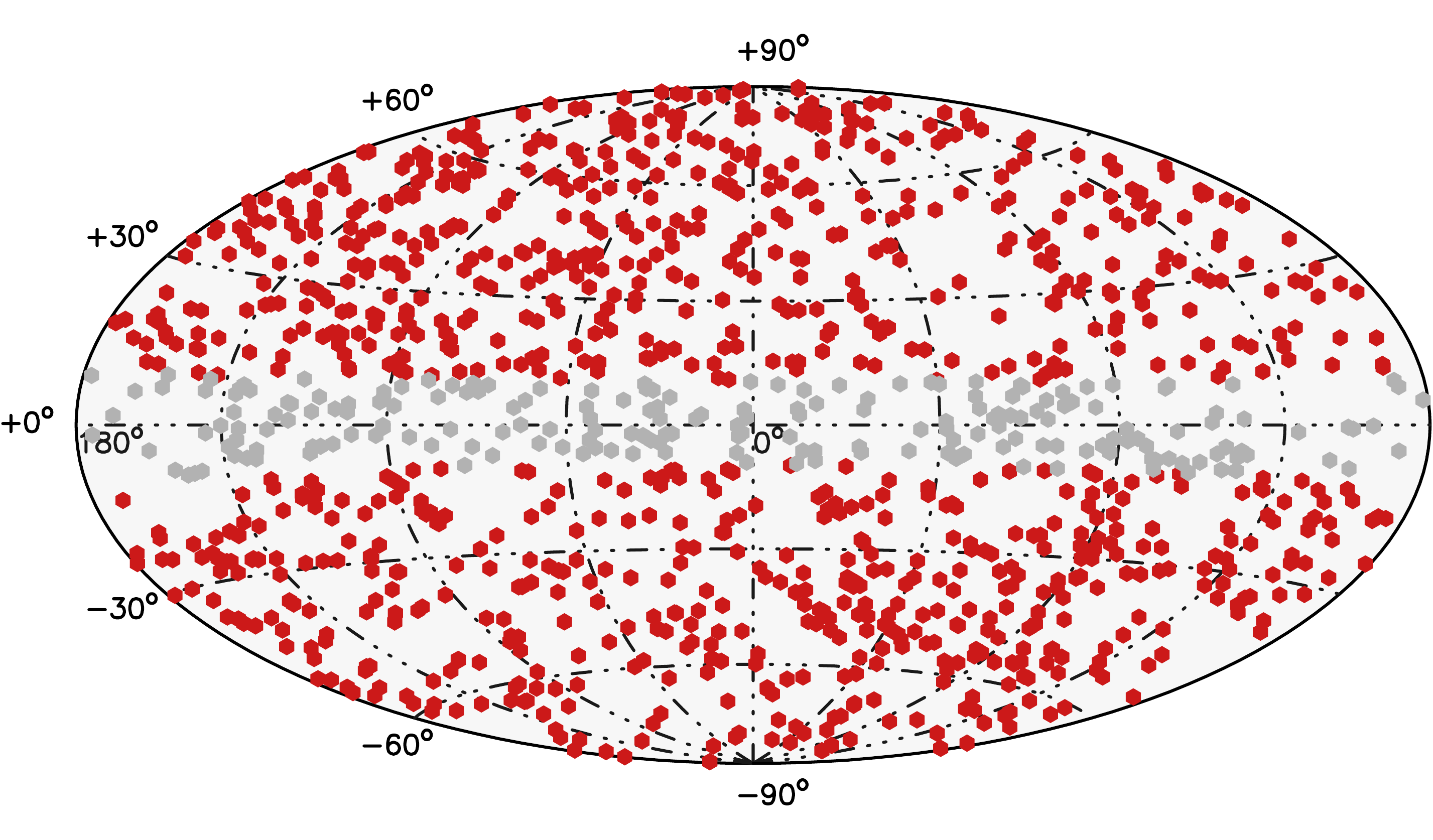}
\caption{Hammer-Aitoff plot in Galactic coordinates of all the Swift-XRT fields centred on GRBs and observed in PC readout mode. The 1,046 fields at Galactic latitude larger then 10 degrees used for the extragalactic survey are shown in red.}
\label{fig:aitof}       
\end{figure}

During the first 15 years of operation, from shortly after launch in late November 2004 till the end of 2019, Swift observed with the XRT telescope over 1,300 GRBs. A Hammer-Aitoff plot of their positions in Galactic coordinates is shown in Fig.\ref{fig:aitof}.

\begin{figure}
\centering
\hspace*{-0.5cm}
\includegraphics[width=9.2cm]{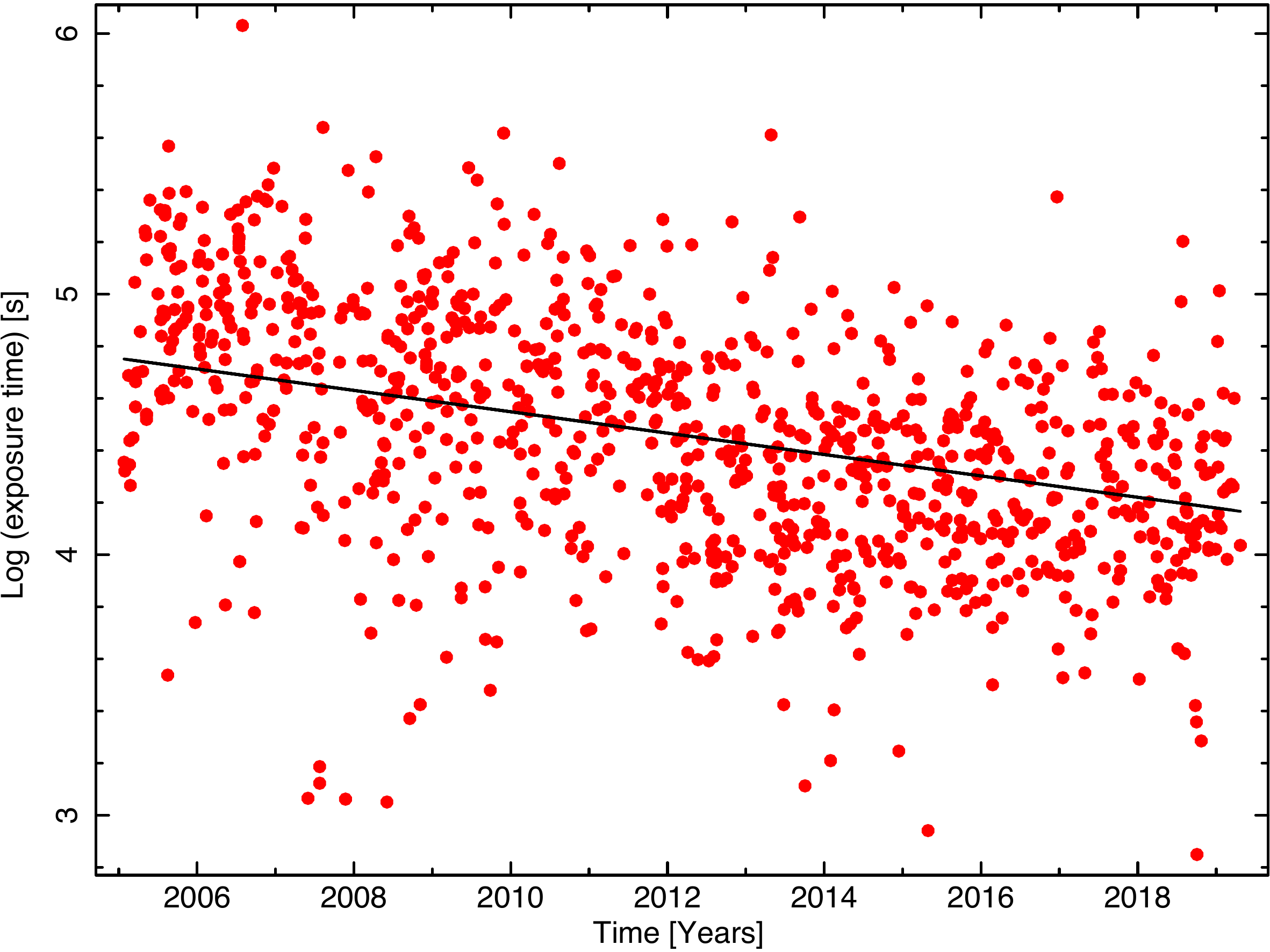}
\caption{The exposure time of the stacked images centred on GRBs as a function of time. A clear trend to lower exposures with time is apparent reflecting the fact that GRBs have been followed for longer times at the beginning of the Swift mission.}
\label{fig:expoVstime}       
\end{figure} 

\begin{figure}
\centering
\hspace*{-0.2cm}
\includegraphics[width=9.9cm]{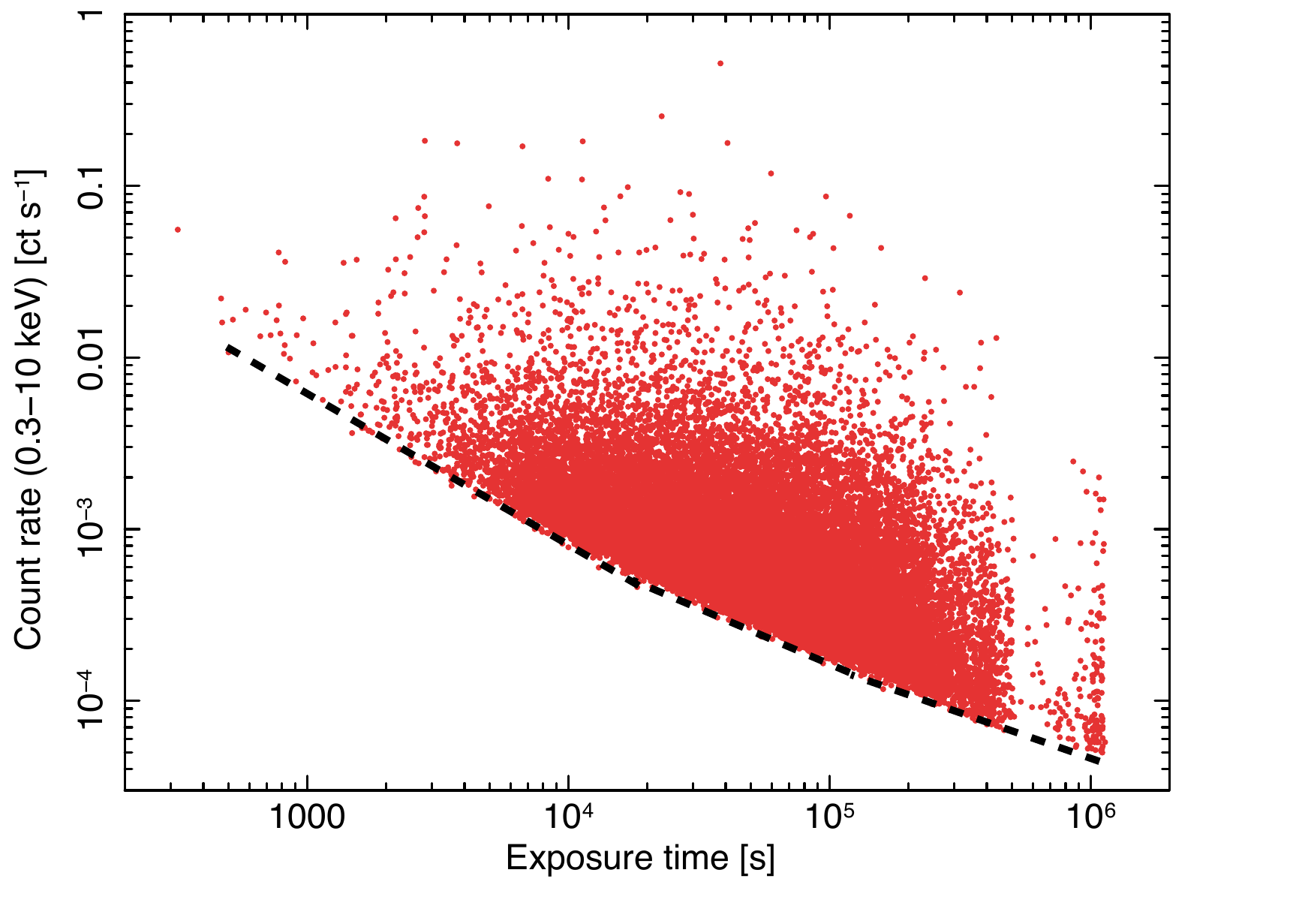}
\caption{Plot of the 0.3-10 keV XRT count-rate versus effective exposure time. The minimum detectable count-rate, which determines the survey limiting sensitivity, is delimited by the black dashed line.
}
\label{fig:sensitivity}       
\end{figure} 

\begin{figure}[ht]
\centering
\includegraphics[width=9.0cm]{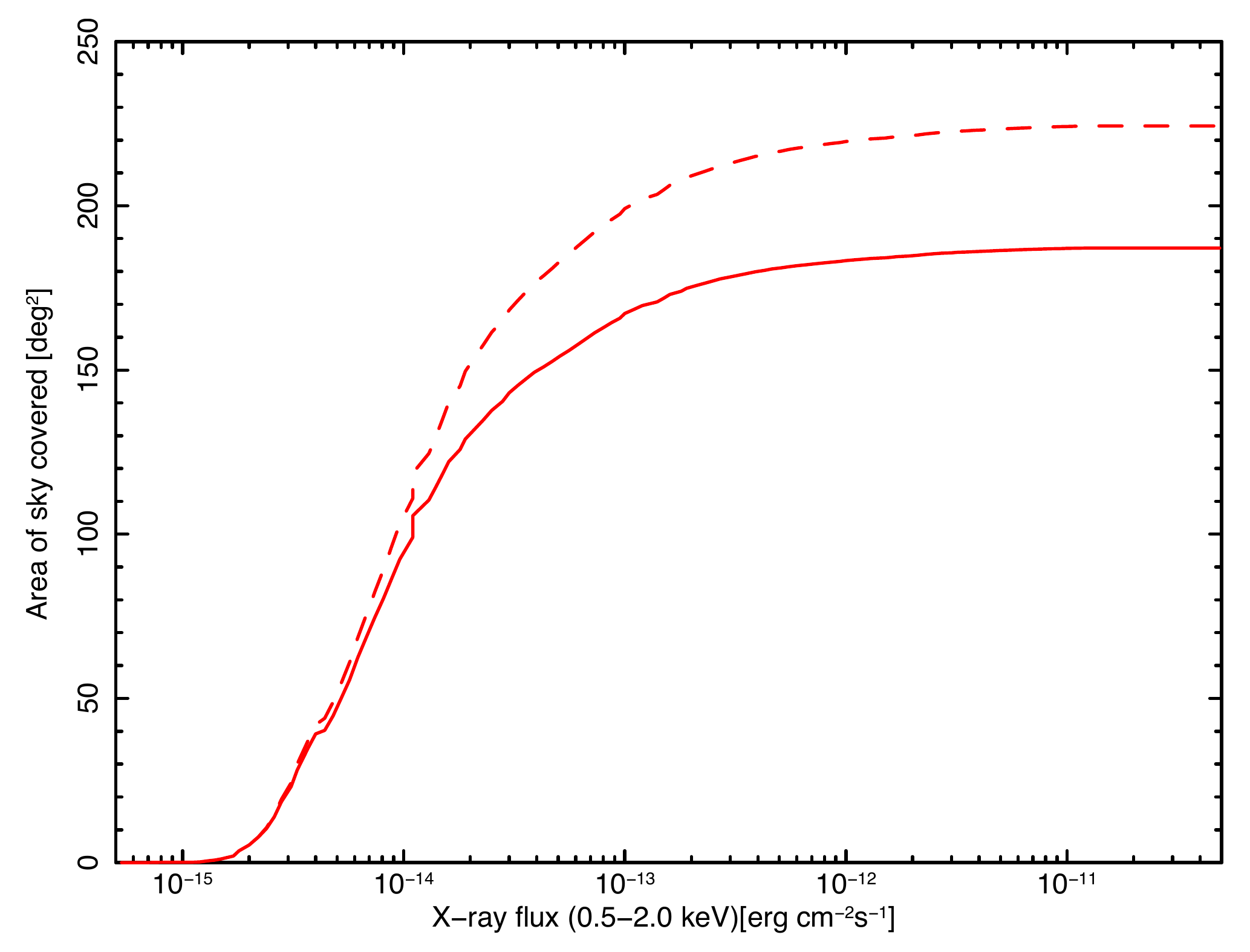}
\caption{The sky coverage of the entire OUSXG survey (dashed line) and that of the clean sample, limited to  high Galactic latitude ($|b| > 10^{\circ}$) XRT fields.}
\label{fig:skycoverage}       
\end{figure} 

\section{XRT data analysis}
Following the approach adopted in Paper I we used the Docker container\foothref{https://www.docker.com} version of the Swift\_deepsky pipeline to analyse all the Swift-XRT observations pointing at GRBs and carried out in Photon Counting (PC) readout mode \citep{xrt,paper1}. 
The Swift\_deepsky software, built on top the official HEASoft data reduction package \foothref{https://heasarc.gsfc.nasa.gov/docs/software/lheasoft}, automatically performs the following tasks:
\begin{itemize}
    \item low-level data and calibration files downloading from one of the official Swift archives
    \item exposure maps and X-ray images generation
    \item stacking of exposure maps and X-ray images
    \item pointlike source detection based on the slide-cell and background determination methods built in the XIMAGE package. In this process the  
    detection threshold is set to a probability of $10^{-4}$ that the photon excess is due to a fluctuation of the background and to a minimum signal-to-noise ratio of 2. These conditions ensure that the expected number of false-positives due to statistics is less than one every 10 fields.
    \item estimation of the count-rates in three energy bands (0.3-1 keV, 1-2 keV and 2-10 keV) based on the XIMAGE/SOSTA tool
    \item spectral parameters estimation based on the count-rates in the three energy bands considered and on the amount of Galactic absorption in the direction of each source.
\end{itemize}

At the end of the processing the software checks the quality of the results flagging fields affected by excessive background or other potential problems \citep[for more details see][]{paper1}.

In most cases Swift pointed at each GRB several times, to follow the evolution of the X-ray flux from the moment of the prompt emission, until the source faded below the XRT sensitivity limit.
Since our goal is to detect faint X-ray serendipitous sources, in order to maximise sensitivity all Swift-XRT observations pointing at the same GRB were stacked into a single X-ray image, resulting in 1,332 summed X-ray fields centred on as many GRBs, 1259 of which passed the quality check mentioned above.

At the beginning of the mission Swift followed most GRBs with long and frequent exposures, in order not to loose any details of the evolution of the X-ray emission. This procedure was later optimised resulting in shorter observations on average. This trend is shown in Fig. \ref{fig:expoVstime} where the exposure time of the stacked images is plotted as a function of time.
All X-ray images considered in this work can be accessed from the Open Universe 
portal\foothref{https://openuniverse.asi.it}, under the "Swift XRT" survey button which is based on the "Aladin Lite" visualiser, developed at CDS, Strasbourg Observatory, France \citep{Aladin2000,Aladin2014}.

\subsection{Sensitivity limits}

The OUSXG survey consists of over one thousand GRB fields characterised by a very wide range of exposure times and sensitivities.
Fig. \ref{fig:sensitivity} illustrates how the minimum detectable count-rate, resulting form the source detection process described in the previous paragraph, changes depending on observation length. The limiting sensitivity is shown by the black dotted line\footnote{ empirically calculated from the data shown in the figure.}, which, for exposures of up to 10,000 seconds, approximates a power law with slope of 1, as in this range the X-ray images are photon limited, that is the cosmic and instrumental backgrounds in the detection area are close to zero. At higher exposures the background level is no longer negligible, and the curve gradually flattens until it reaches the slope of 0.5 at $\sim 5\times 10^{4}~s$  where the survey starts to be fully background limited.

To properly take into account the XRT sensitivity dependence in different parts of the field of view and the non perfectly overlapping images, we add the exposure maps of the single pointings and we divide the resulting stacked map into 1,600 sub images, 24x24 arc-seconds in size, roughly matching the size of the XRT Point Spread Function. The limiting sensitivity of each sub-image is then estimated from the local minimum detectable count-rate (as described above), converted to 0.5-2.0 keV X-ray flux assuming a power law spectrum with energy index of 0.9, absorbed by the amount of Galactic Hydrogen column (NH) in the pointing direction.
The overall sky coverage, that is the total area of sky covered at any given sensitivity, is obtained by summing the contributions of all sub-images of all the GRB fields of the OUSXG Survey. Fig. \ref{fig:skycoverage} plots the sky coverage for the cases of the full survey and for the sub-sample of high Galactic latitude ($|b| > 10^{\circ}$) fields.

\section{The sample of serendipitous X-ray sources}

The Swift\_deepsky pipeline was run on all the 1,332 stacked XRT fields using two Open Universe medium-sized Linux machines located in Rome (ASI) and Pescara (ICRANet). The processing was completely unsupervised and lasted less than two days. The set of serendipitous point-like X-ray sources that were detected in this process and passed the automatic data cleaning procedure described in Paper I includes 30,952 objects, 27,568 of which are located at high Galactic latitudes ($|b| > 10^{\circ}$). 
The effective exposure time at the position of each source is calculated taking into account the image exposure time, vignetting correction, and CCD dead pixels and dead rows. This is obtained by stacking the exposure maps of every pointing that contributes to the stacked X-ray image.
This sample, called OUSXG, once the GRBs target of the observations are removed, is a flux limited unbiased survey of the X-ray sky. 

\begin{figure}[ht]
\centering
\hspace*{-0.5cm}
\includegraphics[width=9.3cm]{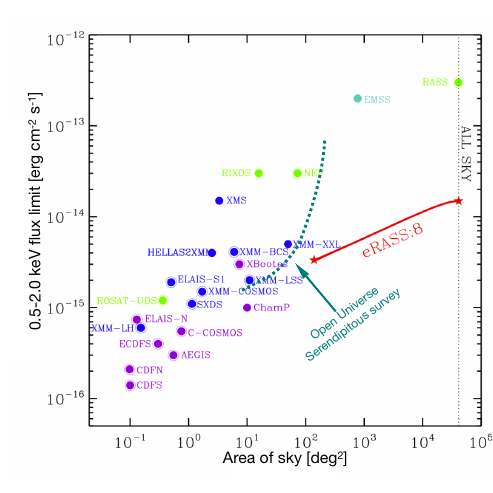}
\caption{The sky coverage of the Swift GRB serendipitous survey (dotted line) compared to all the major existing and upcoming (eROSITA) surveys of X-ray point-like sources.
Adapted from \cite{e-rositasciencebook}}
\label{fig:Surveys}       
\end{figure} 

\begin{figure}[ht]
\centering
\hspace*{-0.2cm}
\includegraphics[width=9.2cm]{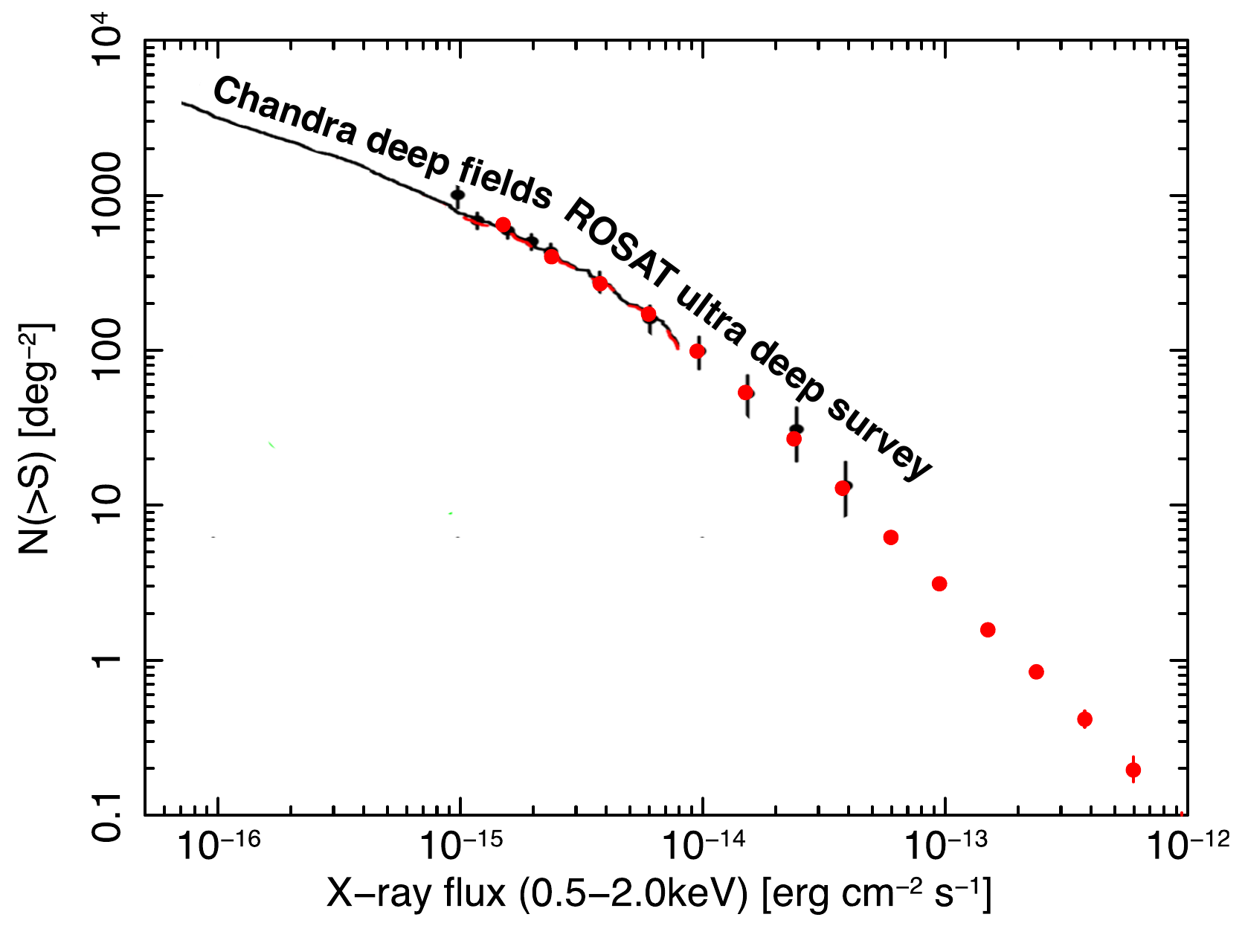}
\caption{A comparison of the 0.5-2.0 keV X-ray LogN-LogS of our Survey (red points) with that measured by  Chandra (solid line) and ROSAT (black filled circles). Adapted from \cite{BH}.}
\label{fig:lognlogsComparison}       
\end{figure}

\subsection{Comparison with other Swift XRT catalogues}

A number of catalogues of serendipitous X-ray sources have appeared in the literature.
Among these we distinguish between general purpose, thematic, and survey catalogues. The first ones  \citep[e.g.][]{DElia2013,Evans2014,2sxps} generally include all X-ray  sources detected by Swift-XRT (or other imaging X-ray telescope) up to a certain date, typically several months before the publication of the list. These are traditional catalogues that reflect the sequence of observations performed by the satellite but do not provide accurate information about the area of sky covered as a function of sensitivity nor the many details that are essential to control the observational biases resulting from the complex scientific and guest observers programs that determine the composition of the archive. 
Thematic catalogues concentrate on a specific type of sources, like for example blazars as in \cite{paper1}, or GRBs etc., while 
survey catalogues \citep[e.g.][and this work]{Puccetti2011,Dai2015} are designed for the purpose of statistical use and therefore pay attention to the need to control observational biases and provide details about the sky coverage as a function of sensitivity. 
So far, all these catalogues have been generated by teams of experts and have been published with an irregular cadence of one every few years.
The approach presented in Paper I, and replicated here in a survey context, constitutes an innovative type of catalogue, that is more dynamical and potentially always up to date, since the Swift\_deepsky software provides the possibility to update an existing catalogue by running the software on new observations by anyone and on most computers at any time.

\subsection{Cross-matching with catalogues of known astronomical sources}

To identify at least a fraction of our serendipitous X-ray sources we have cross-matched the OUSXG sample with several astronomical source lists using a matching radius of 10 arc-seconds for the case of catalogues of point-like objects and 90 arc-seconds for clusters of galaxies. 
The largest table used is the "million quasars" catalogue\foothref{http://quasars.org/milliquas.htm} 
\citep[Milliquas, version 6.4][]{milliquas}, which includes nearly two million AGNs. This resulted in over 6,000 matches (or about 20\% of the total), including sources with redshift up to 5.6.
Since we are interested in finding blazars, we have also cross-correlated our sample with the Open Universe list of blazars, which combines the 5BZCAT \citep{Massaro2015}, the 3HSP \citep{3HSP}, and the 4LAC \cite{4FGL} catalogues, and 
is the largest table of known blazars, obtaining only 34 matching sources. In order to identify blazars that are still uncatalogued we have cross-matched the sample with tables  of radio sources such as the NVSS \citep{NVSS} or the SUMSS21 \citep{SUMSS} catalogues, resulting in nearly 900 matches.  Tab.\ref{tab:cross1} summarises the results of the cross-matching with some of the main catalogues of known astronomical sources,
while Tab. \ref{tab:cross2} gives the number of OUSXG sources that are common to other recent X-ray catalogues.  The choice of a 10 arc-seconds matching radius is somewhat larger than the typical positional error of XRT serendipitous sources. 
We have chosen this value to take into account the positional uncertainties of the other catalogues, that can be up to a few arc-seconds. Given the density of some of the catalogues this could lead to a number of false positive matches. By using the technique of coordinates shifting of one of the matching tables, we estimate that this problem is limited to $\sim$1\% or less.
\subsection{Comparison with other X-ray surveys}

In this section we compare the OUSXG sample to a number of existing or upcoming X-ray surveys.
Fig. \ref{fig:Surveys}, adapted from \cite{e-rositasciencebook}, plots the sensitivity of the most important existing or upcoming X-ray surveys as a function of the area of sky covered. The deepest fields, 
obtained investing several mega-seconds of exposure time
of the largest operating X-ray observatories like Chandra and XMM, reach sensitivities well below $\sim 10^{-15}$ \ergs\ but cover very small areas of sky, whereas the Rosat All Sky Survey \citep[RASS,][]{RASS}, still the only available all sky survey, is relatively shallow, only reaching a flux limit of a few times $10^{-13}$ \ergs in the 0.5-2.0 keV band.
Below $\sim 10^{-13}$ \ergs\ OUSXG (represented by the dashed green line) is currently the largest survey available. Even when the e-Rosita all sky survey \citep{e-rositasciencebook} will be completed, at fluxes of a few times  $10^{-15}$ \ergs\ OUSXG will still be complementary to it, since in this sensitivity regime e-Rosita will only cover a relatively small fraction of sky near the ecliptic poles, whereas the fields of OUSXG 
are located in all parts of the sky.

\subsection{The X-ray LogN-LogS of extragalactic point-like sources}

We have used the OUSXG sample, limited to $|b| > 10^{\circ}$, and the corresponding sky coverage shown in Fig. \ref{fig:skycoverage} to build the (integral) X-ray LogN-LogS of extragalactic sources down to a flux limit of $1.5\times 10^{-15}$ \ergs. Below this flux the survey only covers a few square degrees of sky and the automatic method used for the source detection starts becoming not completely reliable. 
The results are plotted as red symbols in Fig. \ref{fig:lognlogsComparison} superposed to the estimations of other X-ray satellites. The data is tabulated in Tab. \ref{tab:lognsx}.
Given the remarkably good agreement with previous findings we can deduce that: 
\begin{itemize}
    \item the subset of serendipitous point-like X-ray sources with flux $> 1.5\times 10^{-15}$ \ergs\, in the 0.5-2.0 keV band (26,217 objects) detected at Galactic latitude larger than 10 degrees can be considered to be a statistically well-defined sample affected by a negligible fraction of spurious sources;
    \item the sky coverage shown in Fig. \ref{fig:skycoverage} is accurate and can be used, together with suitable sub-samples, to investigate the statistical properties of specific populations of X-ray sources.  
\end{itemize}

\subsection{Serendipitous blazars}

In this paper we exploit the nearly two orders of magnitude better X-ray sensitivity of the OUSXG survey compared to the RASS to correct the incompleteness of the largest existing table of HBL blazars, the 3HSP sample \citep{3HSP}.

In the following we limit ourselves to build a deep radio flux-density limited sample of HBL blazars and use it to estimate its radio LogN-LogS. A detailed study of the statistical properties (in particular the amount of cosmological evolution) of HBL blazars is the subject of a parallel paper (Chang et al. 2020, in preparation) where the sample selected here will be combined with the larger (but incomplete at faint radio flux densities) sample compiled by \cite{3HSP} using multi-frequency data and the RASS survey \citep{RASS}. 

\begin{table}[h]
\begin{center}
\caption{Results of the cross-matching of the clean sample with catalogues of known objects}
\begin{tabular}{llrc}
\hline
Catalogue name & type & no. of&matching \\ 
& & matches&radius ($''$) \\ 
\hline
\hline
Milliquas V6.4\tablefootnote{\cite{milliquas}} & QSOs & 6,038 &10\\
5BZCAT\tablefootnote{\cite{Massaro2015}}, 3HSP\tablefootnote{\cite{3HSP}} &  &  & \\
4LAC\tablefootnote{\cite{4FGL}} & Blazars & 34 &10\\
ASCC\tablefootnote{\cite{ASCC}}& Stars & 676 &10 \\
Principal Galaxies & & &\\
Catalog PGC\tablefootnote{\cite{PGC2003}} & Galaxies & 25&10\\
NVSS\tablefootnote{\cite{NVSS}}, SUMMS21\tablefootnote{\cite{SUMSS}} & Radio sources & 889 &10\\
Zwicky, Abell\tablefootnote{\cite{Zwicky1968,Abell1989}} & Clusters & &\\
Plancksz, SWXCS\tablefootnote{\cite{PlanckSZ,Liu2015}} &  of galaxies & $\sim$140 &90\\
\hline
\hline
\label{tab:cross1}
\end{tabular}
\end{center}
\end{table}

\begin{table}[h]
\begin{center}
\caption{Cross-matching between the clean sample and recent catalogues of X-ray sources}
\begin{tabular}{lrc}
\hline
Catalogue name & number of & Percentage of \\ 
& matches & common sources\\ 
\hline
\hline
2SXPS\tablefootnote{\cite{2sxps}} 
& 25,124 & 81.2 \\
4XMM-DR9\tablefootnote{\cite{4xmm}} 
&  2,001 & 6.5\\
Chandra CSC2\tablefootnote{\cite{csc2}}& 1,361& 4.4 \\
SACS\tablefootnote{\cite{Dai2015}} & 15,667 & 51 \\
\hline
\hline
\label{tab:cross2}
\end{tabular}
\end{center}
\end{table}

\begin{figure}[ht]
\centering
\hspace*{-0.5cm}
\includegraphics[width=9.4cm]{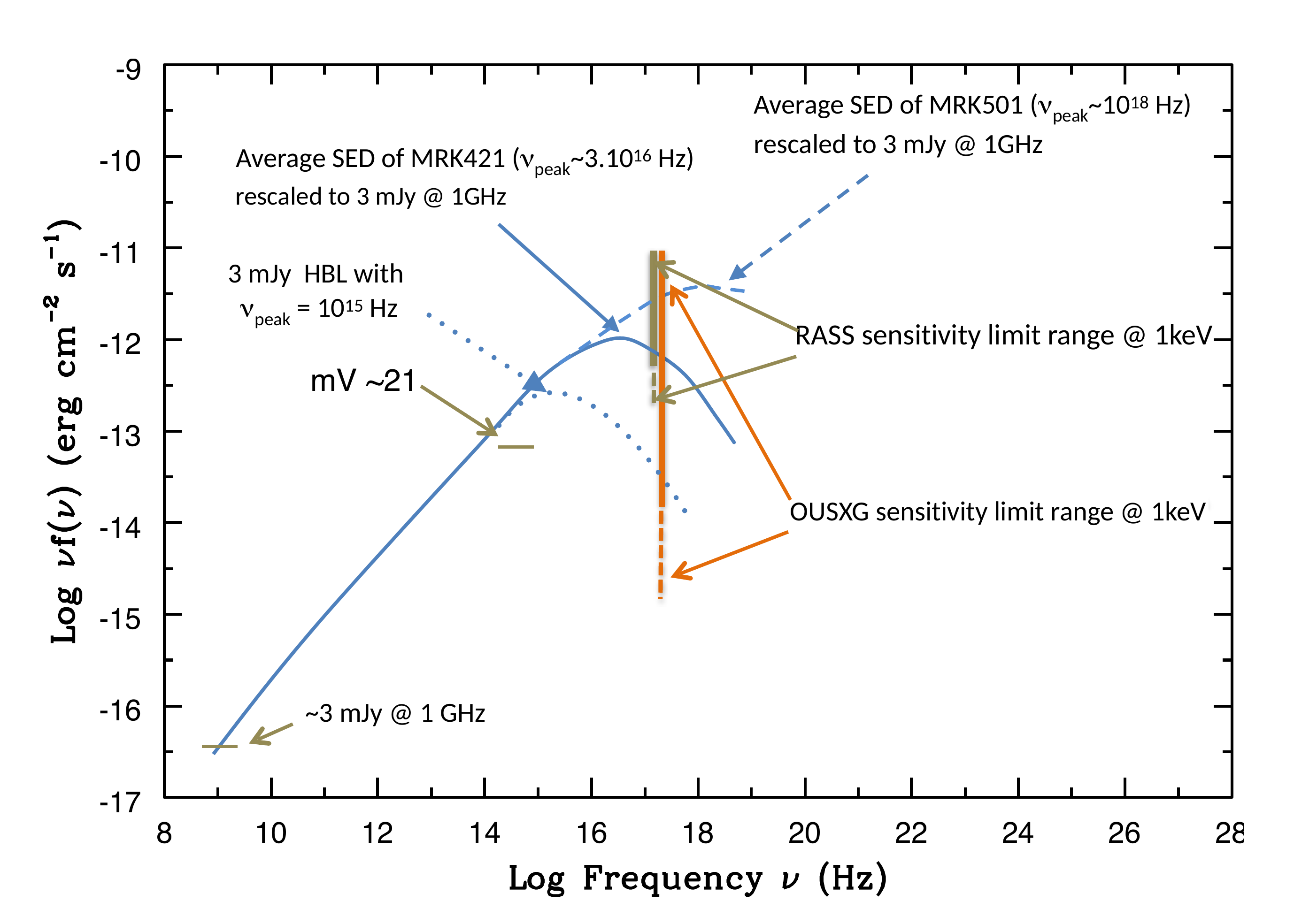}
\caption{Simple scheme comparing the sensitivity ranges of the RASS and the OUSXG surveys at $\sim$ 1 keV to the SED of different types of HBL blazars with radio flux density of 3mJy, close to the sensitivity limit of existing radio surveys. Blazars of this radio flux-density and \nup $\lsim 10^{16}$ Hz can only be detected in the OUSXG survey and are therefore missing in catalogues that require detection in the RASS.}
\label{fig:Sensitivity}       
\end{figure} 
\begin{table*}[ht]
\begin{center}
\caption{The list of HBL blazars in the complete radio flux-density limited sample. 
}
\begin{tabular}{lcccccc}
\hline
Source name &Blazar name & Detected in & Radio & Redshift  & X-ray flux& Log(\nup)\\ 
  && RASS survey &flux d. & & 0.5-2.0 keV& \\
  &&& (mJy) & & (\ergs) & \\
  (1) & (2) & (3) & (4) & (5) & (6) & (7)\\
\hline
\hline
OUSXGJ005224+5003.5 & -- & no &9.1  & 0.53 &2.97E-14& 15.9 \\
OUSXGJ005434$-$3846.2 & -- & no & 22.3 & 0.56? & 4.11E-14& 15.4 \\
OUSXGJ010436+3401.3$^{a,b}$ & -- & no & 3.6 & 0.49 & 1.65E-14 & 15.9\\
OUSXGJ015946+0900.0 & 3HSPJ015945.1+090002 & no & 9.9 & 0.63 & 3.78E-14& 15.9 \\
 OUSXGJ021217$-$0221.9 &3HSPJ021216.9$-$022155 & yes & 25.3 &0.250& 2.04E-12& 17.1 \\
OUSXGJ022540$-$1900.5 &3HSPJ022539.7$-$190035 & no & 5. & 0.40 & 1.10E-13&16.2  \\
OUSXGJ033405$-$3956.3 & -- & no & 4.2 & 0.31 & 2.95E-14 &  15.8\\
OUSXGJ035025+2709.9 & -- & no &3.8 & 0.62 & 1.88E-14& 15.9 \\
OUSXGJ074420$-$6211.0 &3HSPJ074419.1$-$621100  & yes & 48.7 & 0.38 & 2.75E-13 & 16.4 \\
OUSXGJ075119$-$0027.8 & -- & no & 26.5 & 0.27 & 1.58E-13 & 16.1 \\
OUSXGJ080057+0732.5 & 3HSPJ080056.5+073235 & no & 8.1 & 0.44 & 1.19E-12 & 15.6 \\
OUSXGJ083251+3300.1 & 3HSPJ083251.5+330011 & yes & 4.5 & 0.672 & 1.17E-12 & $\gsim 18.0$\\
OUSXGJ085543+1103.2 & 3HSPJ085542.8+110315 & yes & 14.8 & 0.300 &6.66E-14  & 15.8 \\
OUSXGJ085607+7118.8 & 3HSPJ085607.3+711851? & no & 19.0 & 0.31 & 1.09E-13 & 16.1 \\
OUSXGJ091652+5238.4 & 3HSPJ091651.9+523828 & yes & 139. & 0.190 & 9.45E-13 & 16.3\\
OUSXGJ093430$-$1721.3 & 3HSPJ093430.1$-$172121 & yes & 29. & 0.250 & 2.24E-12 & 16.3\\
OUSXGJ111803$-$1531.0 & -- & no & 6.5 & 0.47 & 1.80E-14 & 15.7\\
OUSXGJ113428$-$0702.1 & -- & no & 4.7 & 0.31 & 3.29E-14 & 15.6 \\
OUSXGJ123205$-$1055.9 & 3HSPJ123205.0$-$105600 & no & 11. & 0.19 & 1.44E-13 & 16.1\\
OUSXGJ124231+7634.2 & 3HSPJ124232.3+763418 & yes & 8.8 & 0.48 & 3.15E-13 & 16.4 \\
OUSXGJ125510+2804.2 & 3HSPJ125509.8+280418 & yes & 1.7 & 0.69 & 1.20E-13 & 16.9 \\ 
OUSXGJ154535$-$0019.4 & 3HSPJ154534.7$-$001928 & yes & 6.6 & 0.60 & 1.30E-14 & 15.7 \\
OUSXGJ215413+0004.3$^a$ & 3HSPJ215412.8+000423 & no & 4.3 & 0.217 & 3.00E-14 & 15.9\\
\hline
\hline
\label{tab:hsp}
\end{tabular}
\end{center}
$^a${\footnotesize Uncertain.} 
$^b${\footnotesize More than one possible optical counterpart.} 
\end{table*}

\section{A complete deep radio flux density limited sample of HBL blazars}

The availability of complete flux limited samples is essential for the investigation of the cosmological properties of HBL blazars, especially below $\sim$20-30 mJy, where they appear to show strong signs of cosmological de-evolution \citep{Maccacaro1984,Sedentary}.

Currently the largest compilation of HBL blazars is the 3HSP catalogue \citep{3HSP}, which includes over 2,000 objects. This  sample, despite its large size, cannot be used as such for statistical investigations because 
the selection biases that led to the compilation of the list cannot be controlled.
This is not true for the subsample of $\sim 1,600$ objects that are detected in the RASS survey where the X-ray selection biases can be taken into account, but only for sources with radio flux density \gsim 20-30 mJy. That is because the RASS survey is not sensitive enough to detect the X-ray emission of faint HBL blazars with radio flux densities close to the limits of current radio surveys (NVSS and SUMSS21: $\sim 2-5$ mJy), unless their \nup\, is very large.
The reason for this incompleteness is illustrated in Fig. \ref{fig:Sensitivity}, where we see that HBL blazars with radio flux density of 3 mJy and \nup\, approximately between $10^{15}$Hz and $10^{16}$Hz are expected to have X-ray
fluxes that are well below the sensitivity limit of the RASS survey (grey vertical line) but well within the reach of the OUSXG survey (orange vertical line).
As shown in Fig. \ref{fig:lognlogsComparison} the OUSXG survey instead is reliable down to X-ray fluxes as faint as $\sim 1.5\times 10^{-15}$ \ergs (0.5-2.0 keV), which corresponds to  $\sim 1\times 10^{-15}$ \ergs in \nufnu\, space at 1 keV.
From Fig. \ref{fig:Sensitivity} we see that the OUSXG survey (orange vertical line) can easily include blazars that are as faint as 3 mJy at radio frequencies and with \nup $\ge 10^{15}$ Hz, therefore allowing the selection of a statistically complete radio flux limited sample of HBL blazars down to flux densities of at least 3 mJy.

We searched for previously unknown HBL blazars in our survey by visual inspection of the SED of each OUSXG source that matches a radio source in the NVSS or SUMSS21 catalogues and with an X-ray to radio flux ratio in the range observed in the sample of known HBL sources \citep{3HSP}.
All SED data were retrieved using both the VOU-Blazar tool \citep{VOU-Blazars} and the SSDC SED builder\foothref{https://tools.ssdc.asi.it/SED/}.  
In order to get objective \nup\, values for our sources we use a newly developed deep-learning estimator called \textit{DNNSed}\foothref{https://github.com/tglauch/DNNSed, Glauch et al. in preparation}. The tool uses the multi-frequency (radio to \gr) data available of each blazar, to estimate \nup\, taking into account that some of the emission may not come from the jet, but rather from the host galaxy or from the so called "blue bump". The algorithm has been trained using all the blazars included in the 5BZCat and 3HSP catalogues \citep{Massaro2015,3HSP} for which a robust \nup\, value is available. 

The input to the \textit{DNNSed} tool is a SED file retrieved by means of the VOU-Blazars SED builder \citep{VOU-Blazars}. This multi-wavelength data-set is then segmented in 33 energy bins from radio to \gr s, depending on the availability of experimental data. For a given SED the median and its variance are calculated for each of the bins and fed into the neural network. 
 From the comparison between the \nup\, values estimated by the tool and those of a set of blazars in a control sample, we verified that our \nup\, values 
 are unbiased and reliable up to \nup\, values of $\sim 5\times10^{17}$Hz, with a typical uncertainty of 0.5 dex.
Finally, a visual inspection was carried out to validate the HBL nature of each candidate. We realise that the uncertainty in the determination of \nup\, and the presence of spectral variability in poorly sampled SEDs might induce a small level of mis-classification. Based on our long experience with blazars SEDs we feel that this potential problem should affect no more than one or two sources in the sample. A more precise estimation of this effect would require detailed simulations, which is clearly beyond the scope of this paper.

The set of HBL blazars selected by us constitutes a radio flux limited sample that is statistically complete down to the flux density limits of NVSS and SUMSS21 surveys. It only includes 23 objects, which is indeed a tiny fraction ($<$0.1\%) of the total number of serendipitous sources in the OUSXG survey, most of which are expected to be radio quiet QSOs \citep{BH}, reflecting the extremely low space density of HBL blazars compared to that of all other types of AGN.

Tab. \ref{tab:hsp} presents the sample of HBLs selected in this work.
Columns 1 and 2 give the OUSXG name of the source and the 3HSP name,
if the source is included in the 3HSP catalogue; Column 3 specifies if the source was detected in the RASS survey; column 4 gives the 
radio flux density at 1.4 GHz (or 0.8 GHz) from the NVSSS (SUMMS21) catalogue; column 5 gives the redshift if available; column 6 gives
the 0.5-2.0 keV flux, and the last column gives the \nup\, as estimated by the \textit{DNNSed} tool.

\subsection{The radio LogN-LogS of HBL blazars}

\begin{figure}[ht]
\centering
\hspace*{-0.5cm}
\includegraphics[width=9.2cm]{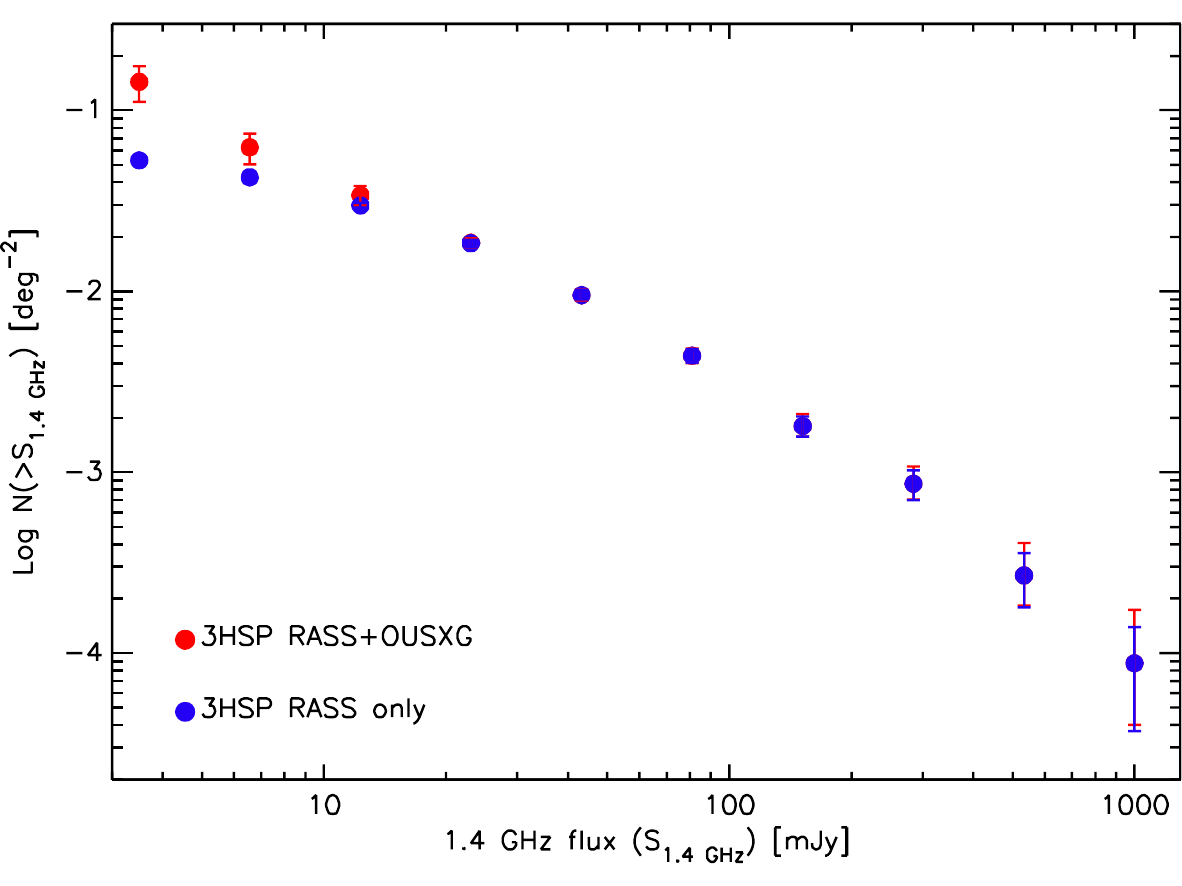}
\caption{The radio LogN-LogS of HBL blazars estimated using the sub-sample 
of 3HSP blazars included in the RASS (blue filled circles) and that of the combined OUSXG and RASS samples (red filled circles). A significant level of incompleteness is clearly present in the 3HSP sample below $\sim 20$ mJy}
\label{fig:lognlogshsp}       
\end{figure}

A detailed analysis of the Cosmological properties of HBL sources will be presented in a dedicated paper (Chang et al. 2020, in preparation). Here we only derive their radio LogN-LogS and compare it to the one presented in \cite{3HSP}, which was based on a sample with some degree of incompleteness at low radio flux density values. 
The radio logN-LogS of our sample of HBLs is shown as red filled circles in Fig. \ref{fig:lognlogshsp} together with that estimated using the sub-sample of 3HSP blazars included in the RASS (blue filled circles) {and the data tabulated in Tab. \ref{tab:lognsradio}}.  As expected, a clear underestimation of the source density in the 3HSP sample is clearly present at radio flux densities below approximately 20 mJy.
\begin{table}[h]
\begin{center}
\caption{Data for the X-ray LogN-LogS shown in Fig. \ref{fig:lognlogsComparison}}
\begin{tabular}{lc}
\hline
0.5-2.0 keV flux & Number density\\
\ergs & deg$^{-2}$\\ 
\hline
\hline
1.50 $\times10^{-15}$ &661. $\pm$ 8.5 \\
2.38 $\times10^{-15}$ &403. $\pm$ 3.0 \\
3.77 $\times10^{-15}$ &268. $\pm$ 2.0 \\
5.97 $\times10^{-15}$ &171. $\pm$ 1.3 \\
9.46 $\times10^{-15}$ &98. $\pm$ 0.9 \\
1.50 $\times10^{-14}$ &53. $\pm$ 0.6 \\
2.38 $\times10^{-14}$ &26.5 $\pm$ 0.42 \\
3.77 $\times10^{-14}$ &12.7 $\pm$ 0.28 \\
5.97 $\times10^{-14}$ & 6.1 $\pm$ 0.19 \\
9.46 $\times10^{-14}$ & 3.1 $\pm$ 0.14 \\
1.50 $\times10^{-13}$& 1.55 $\pm$ 0.10  \\
2.38 $\times10^{-13}$& 0.82 $\pm$ 0.07  \\
3.77 $\times10^{-13}$& 0.40 $\pm$ 0.054  \\
5.97 $\times10^{-13}$& 0.19 $\pm$ 0.042  \\
\hline
\hline
\label{tab:lognsx}
\end{tabular}
\end{center}
\end{table}

\begin{table}[h]
\begin{center}
\caption{Data for the radio LogN-LogS of HBL blazars shown in Fig. \ref{fig:lognlogshsp}}
\begin{tabular}{lcc}
\hline
Flux density& Number density& Number density\\
 & 3HSP only RASS & 3HSP-S\\
mJy& deg$^{-2}$& deg$^{-2}$\\ 
\hline
\hline
3.50 & $5.27\pm0.14 \times10^{-2}$ & $1.43\pm0.32 \times10^{-1}$ \\
6.56 & $4.24\pm0.12 \times10^{-2}$ & $6.22\pm1.21 \times10^{-2}$ \\
12.30 & $2.97\pm0.10 \times10^{-2}$ & $3.38\pm0.41 \times10^{-2}$ \\
23.05 & $1.84\pm0.08 \times10^{-2}$ & $1.85\pm0.10 \times10^{-2}$ \\
43.21 & $9.48\pm0.56 \times10^{-3}$ & $9.48\pm0.66 \times10^{-3}$ \\
81.00 & $4.37\pm0.37 \times10^{-3}$ & $4.37\pm0.43 \times10^{-3}$ \\
151.83 & $1.77\pm0.23 \times10^{-3}$ & $1.77\pm0.29 \times10^{-3}$ \\
284.60 & $8.63\pm1.60 \times10^{-4}$ & $8.63\pm2.15 \times10^{-4}$ \\
533.48 & $2.68\pm 0.90\times10^{-4}$ & $2.68\pm1.38 \times10^{-4}$ \\
1000.00 & $8.78\pm5.07 \times10^{-5}$ & $8.78\pm8.54 \times10^{-5}$ \\
\hline
\hline
\label{tab:lognsradio}
\end{tabular}
\end{center}
\end{table}
\section{Data products availability}

As in the case of Paper I, all the data products generated by this work comply with the principles of transparency put forward by the Open Universe initiative, and are available as high-transparency digital data products in different formats and in a variety of on-line services. In particular:

\begin{itemize}
\item All X-ray images in HIPS format are available from the Open Universe portal under the "Swift XRT" button, via the CDS Aladin visualiser \citep{Aladin2000,Aladin2014}.
\item The sky coverage of the survey and the catalogue giving the same parameters as in Paper I, which include positions, integrated count-rates and fluxes in different energy bands, spectral slopes, and four SED points,  can be retrieved from the Open Universe portal
\foothref{https://openuniverse.asi.it/OU4blazars/Skycoverage0520Fullsurvey.txt}
\foothref{https://openuniverse.asi.it/OU4blazars/Skycoverage0520HighBii.txt}
\foothref{https://openuniverse.asi.it/OU4Blazars/ousxg.fits}.
\item The OUSXG sample can be queried using the on-line query interface at 
BSDC\foothref{http://vo.bsdc.icranet.org} and via VO services.
\item The spectral data have been integrated into the VOU-Blazars/VOU-SED tool, and in the SSDC SED 
builder\foothref{https://tools.ssdc.asi.it/SED/}. 
\end{itemize}

\section{Conclusion}\label{conclusion}

As part of the activities of the Open Universe initiative we have used the Swift\_deepsky Docker pipeline to process all the X-ray images pointing at GRBs generated by Swift-XRT over the last 15 years, from  launch to the end of 2019. Our results can be summarised as follows: 

\begin{itemize}

\item The use of the Swift\_deepsky pipeline, which does not require any expertise in X-ray astronomy, associated with effective cleaning algorithms, allowed us to build a sample of serendipitous X-ray sources that is sufficiently clean to be used for statistical purposes without any visual or manual intervention.
\item Using an X-ray flux limited sub-sample of approximately 26,000 high Galactic latitude sources, we have calculated the X-ray LogN-LogS of extragalactic sources, and showed that it is in excellent agreement with previous measurements.
\item We have built a deep radio-flux limited sample of 23 HBL blazars 
that, combined with larger (and brighter) samples selected using the RASS survey, is suitable for detailed statistical analyses to be presented in a future paper (Chang et al. 2020, in preparation).
\item The radio LogN-LogS of our sample of HBL blazars shown in Fig. \ref{fig:lognlogshsp} clearly implies that the 3HSP sample (and likely all previous samples) of radio faint (f$_{\rm r}$ $\lsim 20$ mJy) HBL blazars suffer from significant incompleteness. 
The cosmological properties of this type of objects estimated with early samples selected in not very deep X-ray surveys might need to be revised based on complete samples that benefit of deep X-ray detections.

\item All data, including the sky coverage of the survey, are available through Open Universe and in other services.
\end{itemize}  

 HBL blazars are the rarest type of AGNs and therefore the selection of sizable samples requires large area relatively deep X-ray surveys. 
 Although OUSXG is much larger than previous similar surveys, it only covers approximately 0.5\% of the sky, and our complete sample of HBL blazars is relatively small, including only 23 objects. 
 Despite their rareness jetted AGN, especially those of the HBL type, are by far the largest population of sources in the high-energy photon and likely neutrino extragalactic sky. Building large samples of these objects is therefore an important contribution to the future of high (MeV-GeV), very-high (GeV-TeV) and extremely-high (PeV and beyond) energy photon and multi-messenger astrophysics.
 Considering all Swift observations, that is not only those centred on GRBs, the area of sky covered by the XRT telescope reaches about 10\% of the sky. 
 The Swift-XRT archive therefore holds the potential for the discovery of about a few hundreds new faint HBL blazars. Although the complex selection biases
 would make this sample hardly usable for detailed statistical studies, it would be very valuable for the reasons mentioned above. 
 The method presented here, which requires the visual inspection of the SEDs of all the candidates, would hardly be applicable in the case of the full Swift archive, as it would require a very large, probably unaffordable amount of manual work. In the future we will search for these and other types of blazars in the Swift and other databases using the machine learning techniques that we are developing building on the experience of this work. 

The methods put forward by Open Universe aimed at simplifying access to space science data reducing or removing the need for specific expertise (in this particular case the Swift\_deepsky Docker container), combined with machine learning techniques, will likely enable future exploitation of large digital archives in accurate and affordable ways by an ever increasing number of cross-discipline researchers, and hopefully in the coming years 
a much larger community of non-professional scientists.   

\noindent {\footnotesize\underline{Disclaimer.} The views expressed herein are those of the authors and do not necessarily reflect the views of the United Nations.}

\begin{acknowledgements}
     
\textbf{PG} acknowledges the support of the Technische Universit\"at M\"unchen - Institute for Advanced Studies, funded by the German Excellence Initiative (and the European Union Seventh Framework Programme under grant agreement no. 291763).

\textbf{CHB} acknowledges the support of ICRANet and the Brazilian government, funded by the CAPES Foundation, Ministry of Education of Brazil under the project BEX 15113-13-2.

\textbf{UBdA} acknowledges the support of a CNPq Productivity Research Grant no. 310827/2016-7 and a Serrapilheira Institute Grant number Serra - 1812-26906. He also acknowledges the receipt of a FAPERJ Young Scientist Fellowship.

\textbf{AVP} and \textbf{OC} are supported by the National Research Council of Argentina (CONICET) by the grant PIP 616, and by the Agencia Nacional de Promoci\'on Cient\'ifica y Tecnol\'ogica (ANPCYT) PICT 140492. \textbf{AVP} and \textbf{OC} are members of the Scientific Research career of the CONICET.

\textbf{NS} acknowledges the support of RA MoESCS Committee of Science, in the frames of the research project No 18T-1C335.

\textbf{RM} acknowledges the financial support of INAF (Istituto Nazionale di Astrofisica), Osservatorio Astronomico di Roma, ASI (Agenzia Spaziale Italiana) under contract to INAF: ASI 2014-049-R.0 dedicated to SSDC.

We thank the anonymous referee for useful suggestions that helped us improving the paper.
\end{acknowledgements}


\bibliographystyle{aa}
\bibliography{1ousxb}

\end{document}